\documentclass[twocolumn,showpacs,preprintnumbers,amsmath,amssymb]{revtex4}

\usepackage{graphicx}% Include figure files
\usepackage{dcolumn}% Align table columns on decimal point
\usepackage{bm}% bold math
\usepackage{epsfig}% Include figure files

\begin{document}

%\preprint{}

\title{Astrophysical relevance of $\gamma$ transition energies}

\author{Thomas Rauscher}%
\affiliation{%
Departement Physik, Universit\"at Basel, CH-4056 Basel, Switzerland}%

\date{\today}

\begin{abstract}
The relevant $\gamma$ energy range is explicitly identified where additional $\gamma$ strength
has to be located for having an impact on astrophysically relevant reactions. It is shown that folding the energy
dependences of the transmission coefficients and the level density leads to maximal contributions for $\gamma$ energies
of $2\leq E_\gamma \leq 4$ unless quantum selection rules allow isolated states to contribute. Under this condition, electric dipole transitions dominate.
These findings allow to more accurately judge the relevance of modifications of the $\gamma$ strength for astrophysics.
\end{abstract}


\maketitle

{\it Introduction.} Predictions made by nuclear theory are essential for all
nucleosynthesis studies but especially for those dealing with explosive
processes proceeding far from the line of nuclear stability. Reactions with
highly unstable nuclei appearing in stellar explosions cannot be directly
studied in the laboratory and most properties required to model the reactions
cannot be measured yet. Current radioactive ion beam facilities are still limited to
a region around stability and often reactions cannot be measured unless either
the target or residual nucleus is long-lived or stable. Because of the low
interaction energies in astrophysically relevant reactions, the statistical
Hauser-Feshbach model \cite{hf} can be used to predict reaction rates provided the
level density in the compound nucleus is sufficiently high \cite{desc}.
The model requires
input based on nuclear structure physics, such as nuclear masses and deformations,
optical model potentials, and nuclear level densities. These are exploited to
calculate transmission coefficients (average widths) which, in turn, determine
the reaction cross section. The reliability of the predictions hinges on two
questions: (i) Is the model applicable for a given reaction and energy, and
(ii) are the relevant inputs known or reliably predicted?

In the past and present, much effort has been and still is devoted to
measure and understand photon strength functions. Often discussed is the
electric dipole (E1) strength which exhibits a pronounced peak at the Giant
Dipole Resonance (GDR) energy. The GDR can be well described by a Lorentzian
although it proved necessary to introduce an additional energy dependence
in the low-energy tail (see, e.g.\ \cite{adndt,ripl-2}).
Recently, a number of experimental investigations
indicated additional strength confined to a small energy range in the low-energy
tail (see, e.g., \cite{alarc87,zilges,gsi,kneisslreview,gsi2}).
Theory provides different possibilities to explain the additional strength, such
as collective vibration of a neutron skin against an inert core composed of
protons and neutrons (pygmy resonance) or other collective modes (see, e.g., \cite{piek06,liang,tso,tso1,vret1,vret2,vretthis}). Accordingly, the
predictions regarding location and width of this additional E1 resonance vary.
The possible impact on astrophysical capture reactions is frequently quoted
as motivation for investigating these phenomena. It is the aim of this paper
to substantiate these claims and to outline some general considerations for
the importance of altered $\gamma$ strengths in astrophysics.

{\it Energetics.}
The average transmission coefficient $T=2\pi \frac{\left<\Gamma \right>}{D}$
is the central quantity in Hauser-Feshbach type calculations, relating an average width to the level
spacing $D$.
The capture cross section $\sigma$ is proportional to the transmission $T_i$ in the initial channel,
to the $\gamma$ transmission $T_\gamma$, and the total transmission $T_\mathrm{tot}$ which comprises
all energetically possible channels:
\begin{equation}
\label{cs}
\sigma (E_\mathrm{proj}) \propto \sum_{J,\pi} \frac{T_i^{U,J,\pi} T_\gamma ^{U,J,\pi}}{T_\mathrm{tot} ^{U,J,\pi}} \quad.
\end{equation}
For laboratory reactions, $T_i$ describes the formation of a compound nucleus from the target
ground state and particle emission back into the initial channel. In astrophysical plasmas, it
additionally accounts for compound formation from thermally excited target states. The formed compound
state is characterized by its excitation energy $U$, spin $J$, and parity $\pi$. The excitation
energy \smash{$U=S_\mathrm{proj}+E_\mathrm{proj}$} is computed from the separation energy of the projectile in the
compound nucleus $S_\mathrm{proj}$ and the projectile energy $E_\mathrm{proj}$.
De-excitation of the compound state by $\gamma$ emission is described by
\begin{eqnarray}
T_\gamma ^{U,J,\pi}& =& \left( \sum_{\mu=0}^{\mu_\mathrm{max}} T^{(U,J,\pi)\rightarrow\mu} _\gamma \right) + \nonumber\\
&+&  \int_{E_{\mu_\mathrm{max}}} ^U \sum_{J',\pi'} \rho (E',J',\pi') T_\gamma ^{(U,J,\pi)\rightarrow (E',J',\pi')}\,dE'\quad,
\label{gamtrans}
\end{eqnarray}
where the first sum includes discrete states $\mu$ and an integration over a level density $\rho=1/D$ is
performed for the energy region with many, unresolved states with energy $E'$, spin $J'$, and parity $\pi'$.
The transmission coefficients on the right-hand side include $\gamma$ emission with all multipole orders allowed by the
standard spin and parity selection rules. Only the lowest multipole orders are important and therefore most Hauser-Feshbach calculations only include E1 and M1 transitions, a few also E2. Here the
descriptions of \cite{rtk97,adndt} were used for E1 and M1. However, the following discussion focuses
on E1 transitions because the results can be understood in terms of the electric dipole alone. For the discussed cases where the level density enters, E1 is dominating. Neglecting M1 totally changes the results only by a few percent. This is in agreement with experiment \cite{m1weak}.

\begin{figure}
\resizebox{0.9\columnwidth}{!}{\rotatebox{0}{\includegraphics[bb=0 150 595 700,clip=]{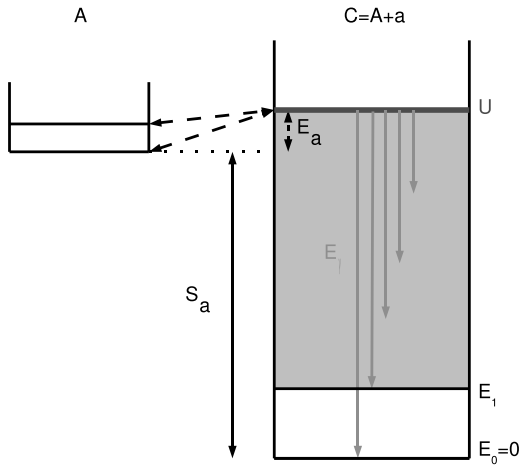}}}
\caption{\label{fig:sketch}Sketch of the relevant energies in a compound capture reaction: target $A$ captures a projectile $a$ by
formation of a compound nucleus $C$. The excitation energy $U$ of the compound nucleus depends on the projectile energy
$E_a$ and the separation energy $S_a$ of the projectile in the compound nucleus. The nucleus $C$ is deexcited via $\gamma$ emission to
discrete states or ``average states'' given by a level density (grey area).
Thus, the possible photon energies $E_\gamma$ can only be in the range $0\leq E_\gamma \leq S_a+E_a$. For astrophysical neutron capture
is $E_a \ll S_a$.
}
\end{figure}

\begin{figure}
\resizebox{0.9\columnwidth}{!}{\rotatebox{270}{\includegraphics[clip=]{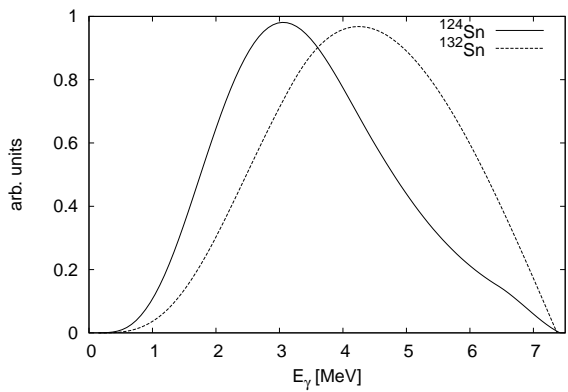}}}
\caption{\label{fig:contrib}The integrand of Eq.\ \ref{gamtrans} in the two compound nuclei $^{124,132}$Sn when
capturing 60 keV neutrons (renormalized to the same maximal value to show the similar shapes).}
\end{figure}

\begin{figure}
\resizebox{0.9\columnwidth}{!}{\rotatebox{270}{\includegraphics[clip=]{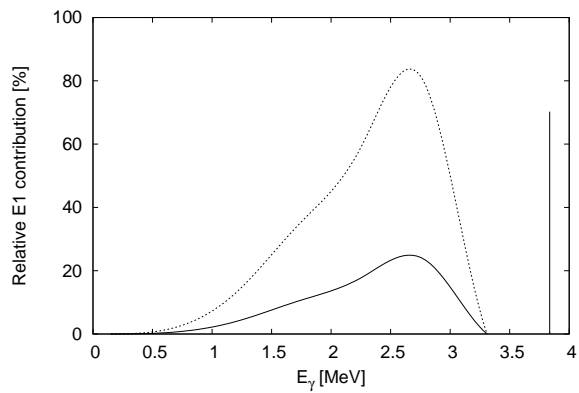}}}
\caption{\label{fig:isol}Relative contribution to $T_\mathrm{E1}$ of Eq.\ \ref{gamtrans} and for $^{135}$Sn(n,$\gamma$) at 60 keV. The full line shows the contributions of the ground state transition and the averaged excited states as given
by Eq.\ \ref{gamtrans} and summed over all $J$, $\pi$ in the compound nucleus. The dotted line shows the relative contribution to the
cross section when including the weighting by $T_i$ as shown in Eq.\ \ref{cs}. In the latter case, the contribution of the ground state transition is completely suppressed because the target nucleus has $J^\pi=7/2^-$ \cite{nndc} and the ground state of the compound nucleus has $J^\pi=0^+$.}
\end{figure}

From the above it follows that the energies of the emitted photons are in the range $0\leq E_\gamma \leq S_\mathrm{proj}+E_\mathrm{proj}$. A sketch of the situation is given in Fig.\ \ref{fig:sketch}.
In astrophysical nucleosynthesis processes, the relevant neutron energies are below
100 keV which is almost negligible compared to neutron separation energies of several MeV, even for
very neutron-rich, unstable nuclei. The situation is different
for charged projectiles because the relevant projectile energies are shifted to $5-10$ MeV, depending on the charges
of target and projectile \cite{ilibook}.

The transmission coefficient $T_\gamma ^{L}$ for $\gamma$ emission with multipolarity $L$ is related to the
(downward) strength function $f$ by $T_\gamma ^{L}=2 \pi E_\gamma ^{2L+1} f(E_\gamma)$.
It is to be noted that only for the strength function $f$ defined in this way there is a direct connection to the
Hauser-Feshbach transmission coefficients.
There are many approaches to derive $f$,
each leading to a basic energy dependence of the E1 transmission given by a Lorentzian
\begin{equation}
\label{e1trans}
T_\mathrm{E1}(E_{\gamma}) \propto \frac{\Gamma_\mathrm{GDR} E^4_\gamma}{(E_\gamma^2 -E^2_\mathrm{GDR})^2 + \Gamma^2_\mathrm{GDR} E^2_\gamma}
\end{equation}
around the Giant Dipole Resonance (GDR) at energy $E_\mathrm{GDR}$ with a width $\Gamma_\mathrm{GDR}$,
but differ in details describing the strength at very low energy \cite{ripl-2}. 

Although transitions with the largest $\gamma$ energies (close to $U$) are the strongest due to the $E_\gamma^4$ dependence (Eq.\ \ref{e1trans}), they receive less weight in the integrand appearing in Eq.\ \ref{gamtrans} because the nuclear level density $\rho$ decreases with increasing $E_\gamma$ with $\rho(E_\gamma)\propto
\exp\left( 2\sqrt{a(U-E_\gamma)}\right)$ (with $a$ being the usual level density parameter). Thus, there will be a competition of few strong transitions
with many weak ones and only a closer inspection of the integrand reveals the relative importance.
As an example, the relative contribution to the integrand of different $E_\gamma$ is shown in Fig.\ \ref{fig:contrib} for two nuclei.
The computation was performed adopting the descriptions of \cite{rtk97,adndt} for $T_\mathrm{E1}$ (and $T_\mathrm{M1}$) and $\rho$, assuming that
$E_\mathrm{proj}=60$ keV and $\mu_\mathrm{max}=0$. It can clearly be seen that the largest contribution stems from the
energy range about mid-way between the ground state and the formed compound state. This conclusion holds even when
employing different descriptions of low-energy GDR strength as given in \cite{ripl-2}. The resulting changes in the
location of the maxima are on a level of 10\% or smaller.

\begin{figure}
\resizebox{0.9\columnwidth}{!}{\rotatebox{270}{\includegraphics[clip=]{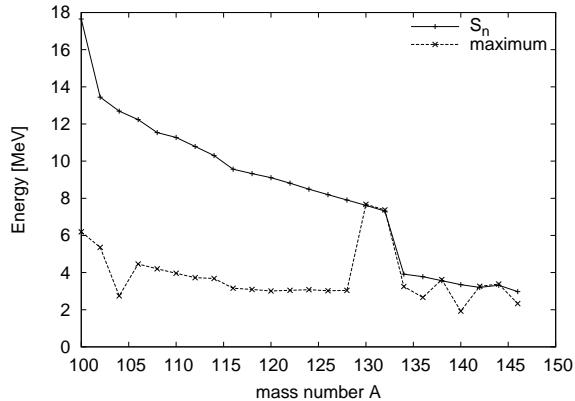}}}
\caption{\label{fig:sn_n}The maximally contributing $\gamma$ energies when capturing 60 keV neutrons on Sn isotopes with odd mass numbers $A$ are compared
to the neutron separation energies $S_\mathrm{n}$ in the compound nuclei. The horizontal axis gives the mass number $A$ of the final (compound) nucleus.}
\end{figure}

\begin{figure}
\resizebox{0.9\columnwidth}{!}{\rotatebox{270}{\includegraphics[clip=]{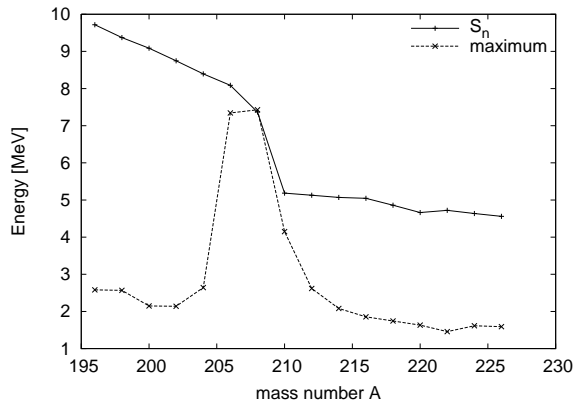}}}
\caption{\label{fig:pb_n}Same as Fig.\ \ref{fig:sn_n} but for Pb isotopes.}
\end{figure}

\begin{figure}
\resizebox{0.9\columnwidth}{!}{\rotatebox{270}{\includegraphics[clip=]{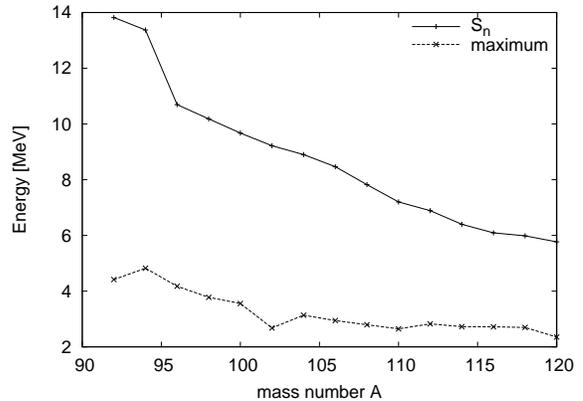}}}
\caption{\label{fig:ru_n}Same as Fig.\ \ref{fig:sn_n} but for Ru isotopes.}
\end{figure}

\begin{figure}
\resizebox{0.9\columnwidth}{!}{\rotatebox{270}{\includegraphics[clip=]{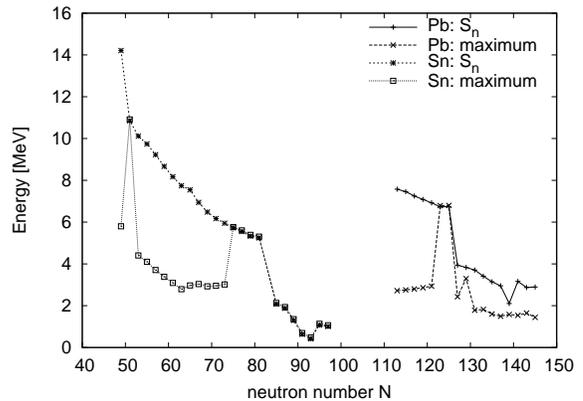}}}
\caption{\label{fig:snpb_n_even}Same as Fig.\ \ref{fig:sn_n} but for neutron capture on isotopes of Sn and Pb with even mass number $A$. The horizontal axis gives the neutron number $N$ of the final nucleus.}
\end{figure}

When the level density is low, the smooth averaged integral is fractionated into contributions of single, isolated states which are accounted for in the first sum of Eq.\ \ref{gamtrans}, including the contribution of the transition to the ground state. For light or closed-shell nuclei, the sum can extend high in excitation
energy. Depending on their spin and parity, these isolated states can carry a considerable fraction of the total strength when they
can be reached by a single E1 transition. Even in the absence of such a fractionation, the contribution of the transitions
to the ground state may not be neglected, i.e.\ the first sum in Eq.\ \ref{gamtrans} has at least one term. However, even when such a state is a main contributor to $T_\mathrm{E1}$ for
a given compound state $(U,J,\pi)$ it may still not be dominating the reaction cross section. That compound state may
not be populated significantly in the reaction due to the spins and parities in the initial channel formed by the
projectile and target. In other words, while $T_\gamma^{U,J,\pi}$ may be dominated by a certain state for a few values of
$J$ and $\pi$, $T_i^{U,J,\pi}$
may be significantly larger for $(J,\pi)$ values where that state is not contributing. Thus, it depends sensitively on the
quantum numbers of projectile, target state, and final state $\mu$ whether $E_\gamma=U-E_\mu$ is a relevant $\gamma$ energy.
An example of this is shown in Fig.\ \ref{fig:isol}, where the ground state contribution would be large but is completely
suppressed. In general, the transitions to the ground state with $E_\gamma=S_\mathrm{proj}+E_\mathrm{proj}$ are only important
in nuclei with an inherent low level density where the application of the statistical model is doubtful, anyway.

{\it Implications for astrophysics.} It has been shown that a change in the low-energy tail of the GDR strength, such as one
caused by a pygmy resonance, can lead to a
significant change in the neutron capture cross sections for neutron-rich nuclei \cite{gor,gorkhan}. Although experiments have indicated
additional E1 strength at low excitation energy \cite{alarc87,zilges,gsi,kneisslreview,gsi2}, the origin of this strength and a prediction of its properties
for more neutron-rich nuclei remains controversial. Nevertheless, using the above arguments the energy can be derived at which the additional strength has to be located to have an impact on astrophysically relevant reactions. As above, the descriptions of \cite{rtk97,adndt} were adopted for $T_\mathrm{E1}$ and $\rho$ and information for ground and excited states was
taken from \cite{nndc} and \cite{mnk}.

\begin{figure}
\resizebox{0.9\columnwidth}{!}{\rotatebox{270}{\includegraphics[clip=]{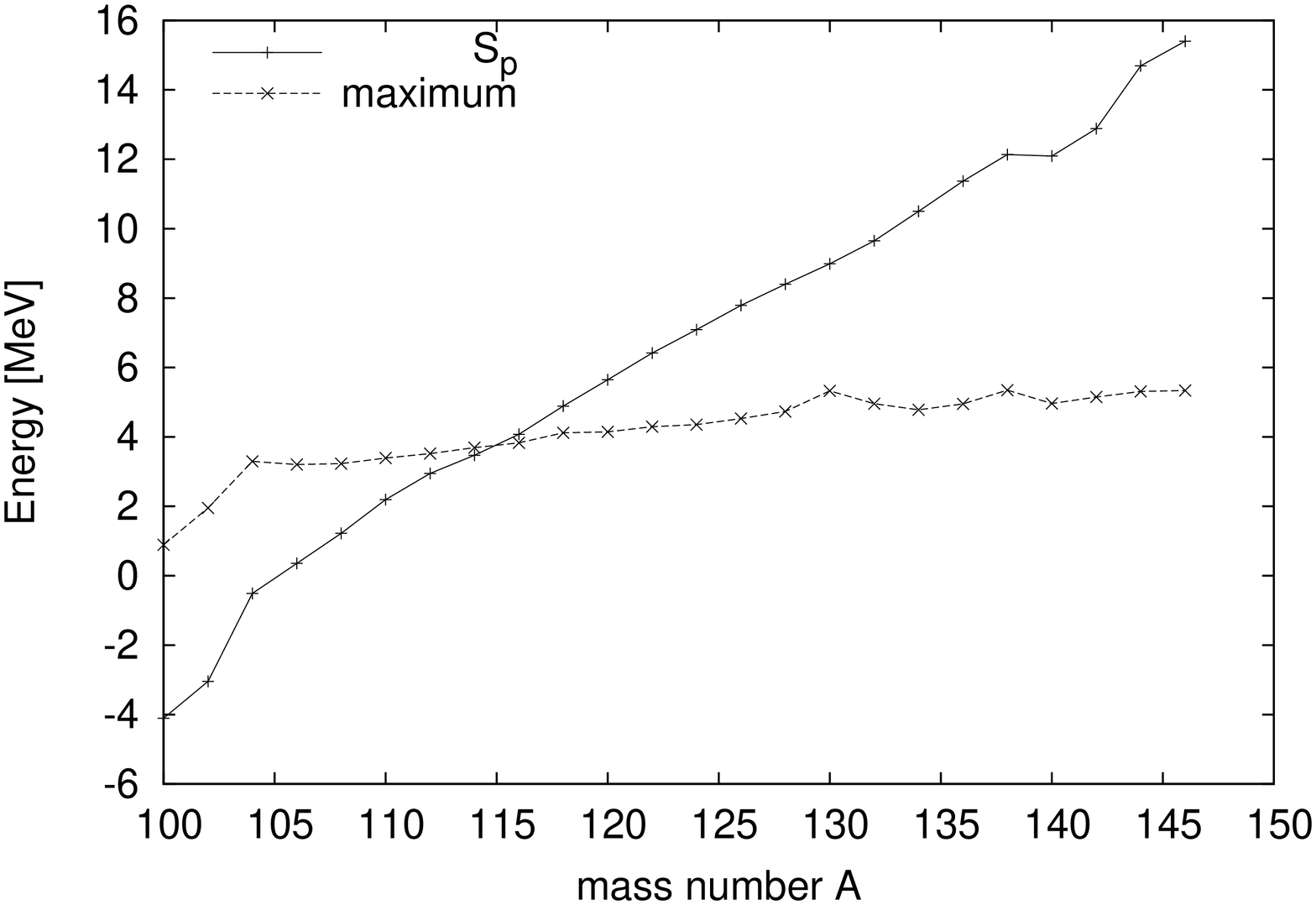}}}
\caption{\label{fig:sn_p}The maximally contributing $\gamma$ energies when capturing 5 MeV protons on Sn isotopes are compared
to the proton separation energies $S_\mathrm{p}$ in the compound nuclei.The horizontal axis gives the mass number $A$ of the final nucleus.}
\end{figure}

\begin{figure}
\resizebox{0.9\columnwidth}{!}{\rotatebox{270}{\includegraphics[clip=]{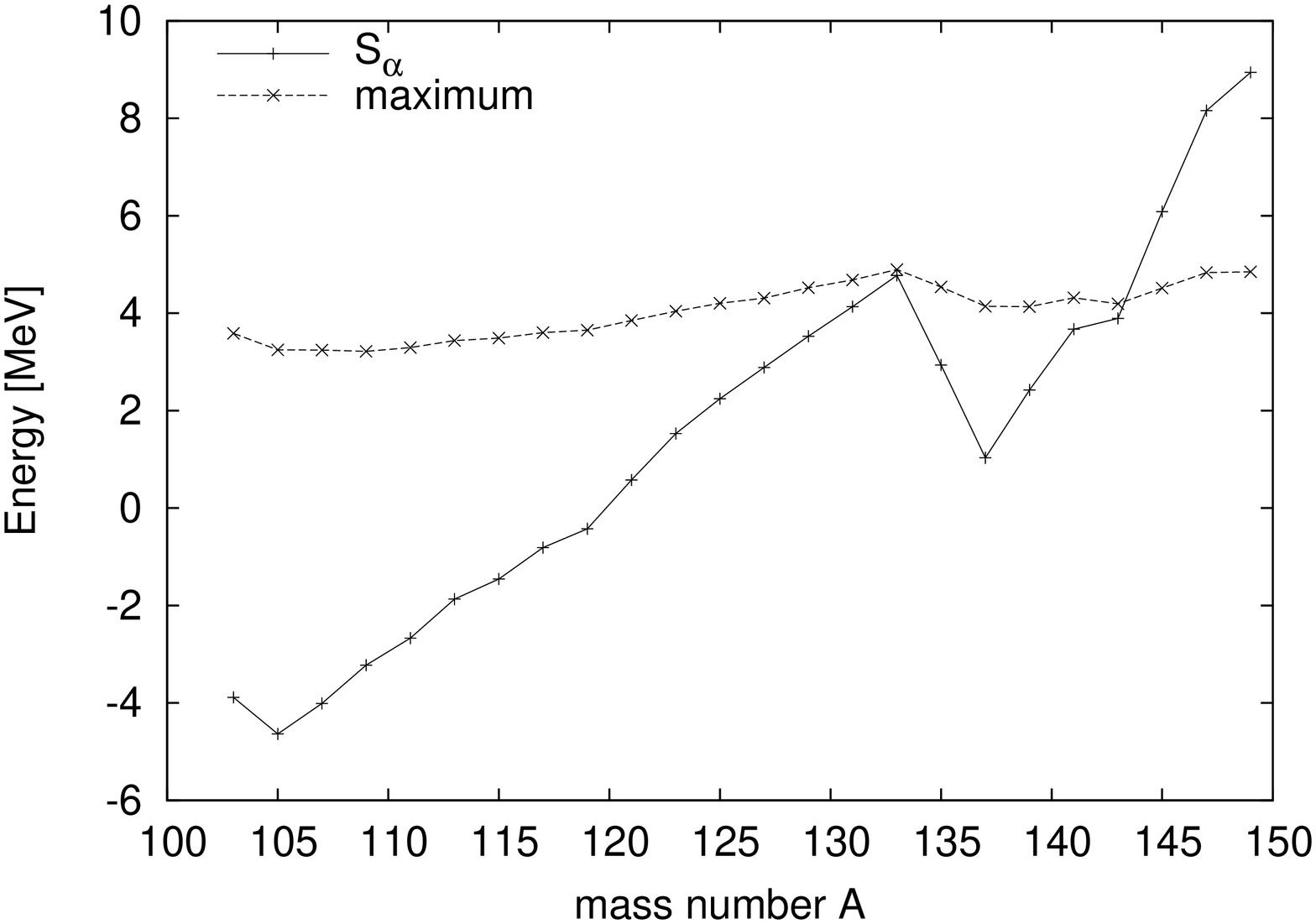}}}
\caption{\label{fig:sn_a}The maximally contributing $\gamma$ energies when capturing 10 MeV $\alpha$ particles on Sn isotopes are compared to the $\alpha$ separation energies $S_\alpha$ in the compound nuclei. The horizontal axis gives the mass number $A$ of the final nucleus.}
\end{figure}

\begin{figure}
\resizebox{0.9\columnwidth}{!}{\rotatebox{270}{\includegraphics[clip=]{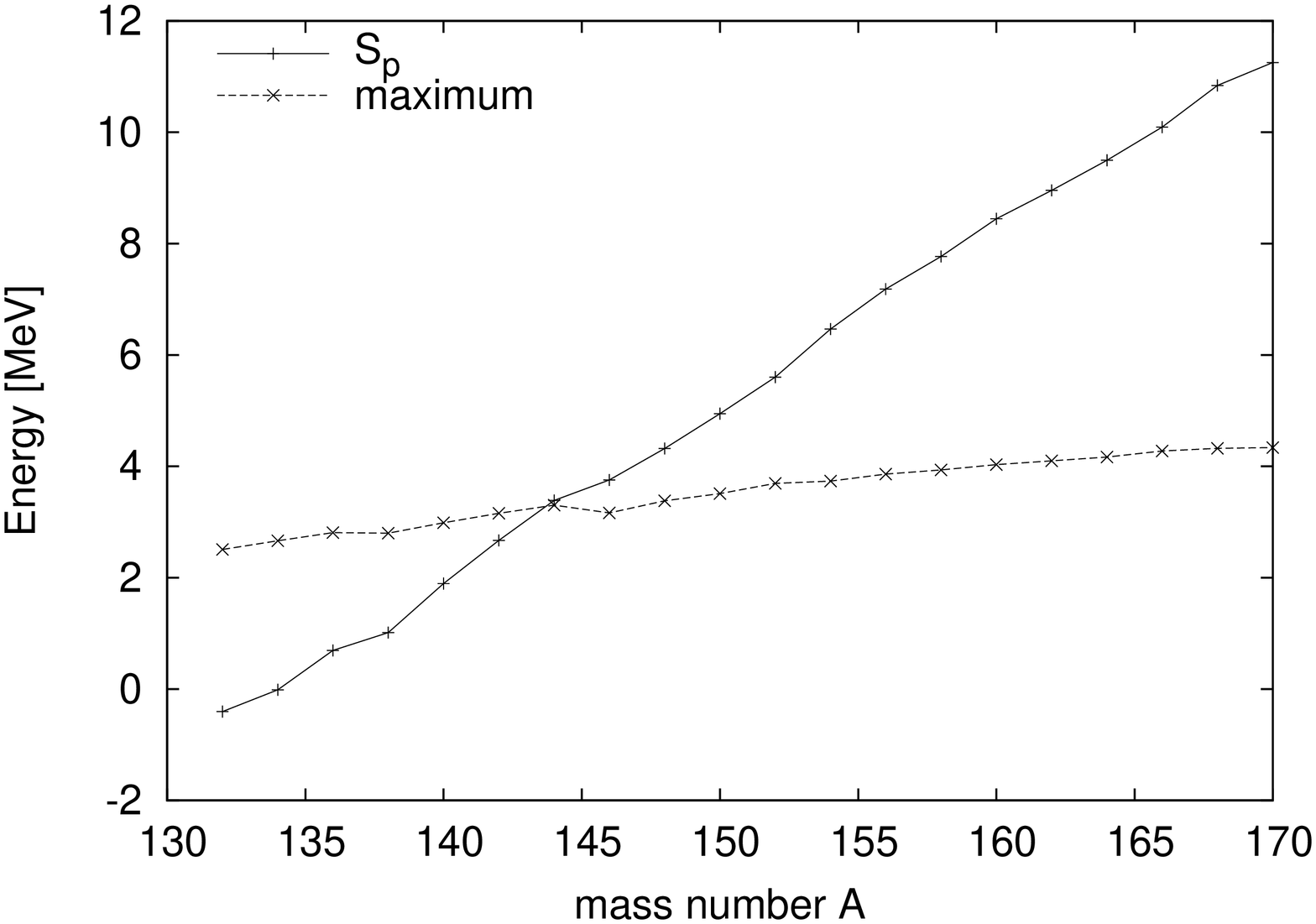}}}
\caption{\label{fig:sm_p}The maximally contributing $\gamma$ energies when capturing 5 MeV protons on Sm isotopes are compared
to the proton separation energies $S_\mathrm{p}$ in the compound nuclei. The horizontal axis gives the mass number $A$ of the final nucleus.}
\end{figure}

\begin{figure}
\resizebox{0.9\columnwidth}{!}{\rotatebox{270}{\includegraphics[clip=]{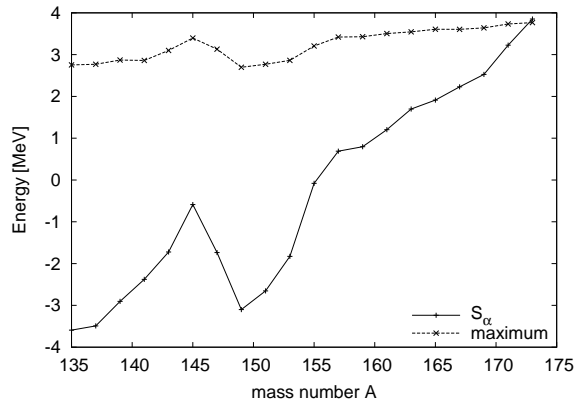}}}
\caption{\label{fig:sm_a}The maximally contributing $\gamma$ energies when capturing 10 MeV $\alpha$ particles on Sm isotopes are compared to the $\alpha$ separation energies $S_\alpha$ in the compound nuclei. The horizontal axis gives the mass number $A$ of the final nucleus.}
\end{figure}

Neutron capture on very neutron-rich nuclei can be relevant in $r$ process nucleosynthesis \cite{prepfriedel,gor}, on proton-rich nuclei in
the $p$ process \cite{arn03,iris08}. Figs.\ \ref{fig:sn_n}$-$\ref{fig:snpb_n_even} show the maximally contributing $\gamma$ energies for sequences of Sn, Pb,
and Ru isotopes, respectively. The weighting by $T_i$ is considered according to Eq.\ \ref{cs}.
The neutron separation energy is decreasing in an isotopic chain with increasing neutron number
but at the same time the maximal level density is also decreasing due to the lower
excitation energies $U$ encountered. In almost all shown cases the energy of the maximal E1 contribution is in the range
$2\leq E_\gamma \leq 4$ MeV. Jumps can be found for nuclei where the selection rules allow the compound-nucleus ground state to be reached by a combination of s-wave neutrons and E1 $\gamma$s. In this case the relevant $E_\gamma$, sensitive to
modifications of the strength $f$, is a sharply defined energy and equal to $S_\mathrm{n}+E_\mathrm{proj}\approx
S_\mathrm{n}$.

Experimentally, additional E1 strength
was found in $^{132}$Sn and attributed to a pygmy resonance \cite{gsi}. However, it is located several MeV above $S_\mathrm{n}$, in accordance
with theoretical predictions. Thus, it does not affect the neutron capture cross section. Comparing to an available prediction of
the pygmy resonance energy within an isotopic chain, it can be seen that it is predicted to always lie well above $S_\mathrm{n}$ \cite{vret1,vret2,vretthis}.
If this is confirmed, it would mean that the pygmy resonance does not play a role in astrophysical neutron capture unless
it is sufficiently wide to bring some additional strength below $S_\mathrm{n}$. To have a wide and strong pygmy resonance may prove difficult, however, without violating the E1 sum rule (Thomas-Reiche-Kuhn sum rule). On a side note it should be remembered that the Hauser-Feshbach model
cannot be applied when the level density is too low at the compound formation energy \cite{rtk97}. This will occur in nuclei with $S_\mathrm{n}$
of only a few MeV. In the absence of resonances direct capture
will dominate neutron capture on nuclei close to the neutron dripline, which does not excite collective modes in the $\gamma$ emission
to the continuum and is not sensitive to pygmy effects. Furthermore, most $r$ process models predict the $r$ process path to be
characterized by an (n,$\gamma$)$-$($\gamma$,n) equilibrium in which individual cross sections do not play a role \cite{chrigel}. In these
models, neutron captures only have a moderate effect during the relatively fast freeze-out at the end of the $r$ process. Those
captures will also occur closer to stability as compared to the $r$ process path.

Energetically, the situation is different for capture of charged particles. Due to the Coulomb barrier, the astrophysically
relevant interaction energies are shifted to higher energies. Reactions with charged particles occur at the line of stability
and on the proton-rich side of the nuclear chart in $p$ \cite{woo78,arn03}, $rp$ \cite{schatz}, and $\nu p$ process 
\cite{carlaprl} nucleosynthesis. In the most extreme case,
non-equilibrium proton captures involve protons of $5-7$ MeV maximum energy. For $\alpha$ particles, the maximum energies are
around $10-12$ MeV. These energies depend on the charge of the target nucleus \cite{ilibook}. Since they are comparable to or
exceed the proton or $\alpha$ separation energies, respectively, the sensitive $E_\gamma$ will be at or above the projectile
separation energy in the compound nucleus. Otherwise, similar rules apply as for neutron capture. However, a widely accepted
explanation of the pygmy resonance is that it is caused by a collective motion of a neutron skin against a proton-neutron core.
This is not likely to occur in neutron-deficient nuclei but other ways to generate additional strength beyond the GDR may
still be found. Figs.\ \ref{fig:sn_p}$-$\ref{fig:sm_a} show the relevant $\gamma$ energies in comparison to the separation
energies for proton and $\alpha$ capture on Sn and Sm isotopes. It is interesting to note that the energies again remain in
the range of about $2\leq E_\gamma \leq 4$, even for $\alpha$ captures with negative $Q$ values.

{\it Summary.} The relevant $\gamma$ energy range was explicitly identified where additional $\gamma$ strength
has to be located for having an impact on astrophysically relevant reactions. It was shown that folding the
energy dependences of the transmission coefficients and the level density leads to maximal contributions for $\gamma$ energies in the range of about $2\leq E_\gamma \leq 4$ MeV. The distributions show a full width at half maximum of about 2 MeV.
Quantum selection rules allow isolated states to contribute only in special cases, mainly for neutron capture around closed shells
or at low neutron separation energy when also the application of the statistical model becomes problematic.
This can be seen in Figs.\ \ref{fig:sn_n}, \ref{fig:pb_n}, and \ref{fig:snpb_n_even} for neutron capture on $^{208-211}$Pb, as well as on $^{100,124,126,128-130}$Sn and most isotopes beyond $^{130}$Sn where the neutron separation energy remains low.
These findings allow to more accurately judge the astrophysical relevance of modifications of the $\gamma$ strength, either found
experimentally or derived theoretically. The importance of experimentally
obtaining spectroscopic information for nuclei with low inherent level densities far off stability is also evident.

{\it Acknowledgments.} This work was supported by the Swiss National Science Foundation (grant 2000-105328).

\end{document}